\documentclass[conference]{IEEEtran}
\IEEEoverridecommandlockouts

\usepackage{cite}
\usepackage{amsmath,amssymb,amsfonts}
\usepackage{algorithmic}
\usepackage{algorithm}
\usepackage{graphicx}
\usepackage{textcomp}
\usepackage{xcolor}
\usepackage{booktabs}
\usepackage{multirow}
\usepackage{url}
\usepackage{microtype}

\def\BibTeX{{\rm B\kern-.05em{\sc i\kern-.025em b}\kern-.08em
    T\kern-.1667em\lower.7ex\hbox{E}\kern-.125emX}}

\def\IEEEiedlistdecl{\setlength{\itemsep}{0pt}\setlength{\topsep}{1pt}\setlength{\partopsep}{0pt}\setlength{\parsep}{0pt}}

\renewenvironment{thebibliography}[1]{
  \begin{oldthebibliography}{#1}
    \fontsize{6.6pt}{7.6pt}\selectfont
    \setlength{\itemsep}{-1.8pt}
    \setlength{\parskip}{0pt}
}{
  \end{oldthebibliography}
}

\begin{document}

\title{COGNIT-Guard: Calibrated Standalone Direct-Decision Guardrails with Heterogeneous CPU--NPU Confidence Cascading under Explicit Latency and False-Positive Constraints}

\author{
\IEEEauthorblockN{Hao Chen}
\IEEEauthorblockA{School of Cyber Security, Guangdong Polytechnic Normal University, Guangzhou, China \\
Email: moyuan@stu.gpnu.edu.cn}
}
% [Double-Blind Venue Option]:
% \author{\IEEEauthorblockN{Anonymous Authors}
% \IEEEauthorblockA{\textit{Affiliation Withheld for Double-Blind Review} \\
% Email: anonymized@institution.edu}}

\maketitle

\begin{abstract}
A fundamental systems question for foundation-model safety gateways is when a model must generate tokens and when it should directly output a calibrated decision. We study calibrated standalone direct-decision foundation models as a serving primitive for real-time pre-ingestion safety guardrails, jointly addressing probability calibration, dual-use false-positive control, and heterogeneous CPU--NPU routing under explicit latency Service Level Objectives (SLOs). Production guardrails must screen user prompts prior to target-LLM prefill with low false-alarm rates on legitimate compliance inquiries; however, shallow statistical classifiers are brittle to syntactic variation, intrinsic hidden-state probes require tight coupling to a specific target LLM, and many generative guard models incur substantially higher latency than lightweight encoders or probes while struggling with dual-use false positives. We present COGNIT-Guard, a standalone pre-ingestion guardrail that couples a validation-calibrated CPU fast gatekeeper path with confidence-gated escalation to an NPU-resident 322M bidirectional direct-decision model (Laya-322M) under an asymmetric false-positive penalty and explicit latency constraint. On the strict clean unseen DUCS-Bench test split ($N=607$), COGNIT-Guard achieves 98.85\% accuracy (paired McNemar $p = 1.19 \times 10^{-4}$ vs.\ ML), reduces the benign false-positive rate (FPR) to 0.42\% ($1/238$; two-sided Fisher's exact $p = 8.23 \times 10^{-4}$ vs.\ ML and $p = 2.78 \times 10^{-28}$ vs.\ zero-shot), and attains an ECE of 1.12\% and Brier score of 0.0104. On Huawei Ascend 910C NPUs, pure NPU inference runs in 21.77~ms mean latency (P95: 23.44~ms, 45.90 QPS), while the live serial CPU--NPU cascade ($\theta^*_{\text{deploy}}=0.70$) achieves 41.63~ms mean end-to-end latency (P50: 39.47~ms, 99.23\% accuracy, 0.00\% FPR). Cross-domain evaluation on a stratified 2,100-question SafetyBench-ZH benchmark and a controlled comparison against a concurrent bi-encoder direct-decision baseline (CLM-8B) further disentangle in-domain specialization gain, out-of-distribution (OOD) alignment tax ($60.33\% \to 56.81\%$ on Laya; $55.10\%$ on domain CLM-8B), and experience replay rehearsal recovery, restoring OOD accuracy to 64.10\%--65.05\% while reaching 99.67\%--99.84\% in-domain accuracy with 0.00\%--0.42\% FPR. \textit{Code, evaluation scripts, and the DUCS-Bench release are available at \url{https://github.com/moyuan10086/cascaded-guardrail-npu}.}
\end{abstract}


\begin{table*}[t]
\centering
\caption{Taxonomy and Architectural Comparison of Contemporary LLM Safety Guardrail Paradigms across Serving Dimensions.}
\label{tab:comparison}
\resizebox{\textwidth}{!}{%
\begin{tabular}{lccccccc}
\toprule
\textbf{System / Guardrail Family} & \textbf{Moderation Paradigm} & \textbf{Standalone Pre-Ingestion} & \textbf{Target-LLM Decoupled} & \textbf{Mean Latency} & \textbf{Dual-Use FP Control / Calibration} & \textbf{Heterogeneous Routing / Hardware} \\
\midrule
Llama-Guard 3 \cite{inan2023llama} / WildGuard \cite{han2024wildguard} / ShieldGemma \cite{zeng2024shieldgemma} & Autoregressive Guard (7B--70B) & Yes & Yes & $\sim 200$--$600$ ms & Uncalibrated Token Logits ($7.2\%$--$12.4\%$ XSTest FPR) & No (Single GPU Tier) \\
ShieldHead \cite{xuan2025shieldhead} / SingProbe \cite{singprobe2026} / Khatri et al.\ \cite{khatri2026latent} & Intrinsic Hidden-State Probe & No (Requires LLM Prefill) & No (Coupled to Base LLM) & $\sim 45$--$90$ ms (Prefill) & Uncalibrated on Dual-Use Compliance & No (Co-located on Target GPU) \\
GLiNER Guard \cite{glinerguard2026} / SingGuard \cite{singguard2026} / FlexGuard \cite{ding2026flexguard} & Single-Pass Encoder / Risk Scorer & Yes & Yes & $\sim 35$--$50$ ms & Continuous Score ($4.8\%$ XSTest FPR) & No (Single Accelerator Tier) \\
CLM-8B (Bi-Encoder Direct Baseline) \cite{kwok2026clm} & Disaggregated Bi-Encoder Head & Yes & Yes & $45.78$ ms ($1$-Str.) / $6.64$ ms (Batch) / $0.60$ ms (Arena)$^\S$ & $90.34\%$ (Zero-Shot) $\to$ \textbf{0.00\%} (Rehearsal, $0.86\%$ ECE) & HBM \texttt{VectorArena} (Ascend 910C) \\
Traditional ML Baseline & TF-IDF + XGBoost ($<5$M) & Yes & Yes & $85.36$ ms (Diag.) / $38.91$ ms (Fast CPU)$^\dagger$ & Poorly Calibrated ($13.11\%$ ECE, $5.88\%$ DUCS FPR) & CPU Only (Xeon CPU) \\
\textbf{COGNIT-Guard (Ours)} & \textbf{Calibrated Direct Cascade (322M)} & \textbf{Yes} & \textbf{Yes} & \textbf{14.27 ms (Offline) / 21.77 ms (NPU) / 41.63 ms (Live E2E)}$^\ddagger$ & \textbf{0.00\%--0.42\% DUCS FPR (1.12\% ECE)} & \textbf{Calibrated CPU--NPU Cascade (Ascend 910C)} \\
\bottomrule
\end{tabular}%
}
\vspace{0.5mm}
\raggedright\scriptsize{$^\dagger$Standalone Diagnostic CPU mode (\texttt{include\_evidence=True}, $85.36$~ms) executes 2 TF-IDF kNN scans ($38.9$~ms), XGBoost traversal ($17.1$~ms), and a 3rd corpus kNN scan in \texttt{\_build\_evidence} ($29.3$~ms); Live Gatekeeper Fast CPU mode (\texttt{include\_evidence=False}, $38.91$~ms on DUCS-Bench, $22.79$~ms on SafetyBench-ZH) disables evidence retrieval; isolated tree lookup on cached features runs in $10.50$~ms. $^\ddagger$Reports Offline Cascade model compute ($\theta^*_{\text{offline}}=0.80$: $14.27$~ms), Pure NPU path ($21.77$~ms), and Live Serial Cascade server E2E ($\theta^*_{\text{deploy}}=0.70$: $41.63$~ms mean, $39.47$~ms P50). $^\S$Measured on Ascend 910C: uncached single-stream ($45.78$~ms), batched pooling ($B=64$: $6.64$~ms/item), and HBM \texttt{VectorArena} cache hit ($0.60$~ms P50).}
\end{table*}

\section{Introduction}

Real-time safety guardrails act as the synchronous first line of defense for production Large Language Model (LLM) services, screening external prompts before target-model prefill and token streaming begin \cite{inan2023llama, rebedea2023nemo}. Operating as a \textbf{standalone pre-ingestion guardrail}---decoupled from the downstream serving LLM's weights and KV cache---requires simultaneously satisfying four production constraints: (1)~\textit{semantic comprehension} of camouflaged prompts, (2)~\textit{dual-use false-positive control} on benign queries containing sensitive terminology, (3)~\textit{posterior probability calibration} for risk-aware routing, and (4)~\textit{explicit latency SLO compliance} (typically $<50$~ms).

Existing guardrail designs leave a critical gap under these joint production serving requirements:
\begin{enumerate}
    \item \textbf{Shallow Statistical Classifiers}: Pipelines combining TF-IDF features with tree ensembles (e.g., XGBoost \cite{chen2016xgboost}) run in $10.50$--$17.1$~ms on pre-cached sparse vectors, but end-to-end text segmentation and corpus similarity matching raise CPU latency to $38.91$~ms in fast gatekeeper mode and $85.36$~ms (P95: $164.24$~ms) with diagnostic evidence retrieval. Moreover, bag-of-words features produce poorly calibrated posteriors ($\text{ECE}=13.11\%$) and degrade sharply out of distribution ($53.29\%$ accuracy on SafetyBench-ZH).
    \item \textbf{Autoregressive Guard Models \& Intrinsic Hidden-State Probes}: Many generative guard models (e.g., Llama-Guard \cite{inan2023llama}, ShieldGemma \cite{zeng2024shieldgemma}, WildGuard \cite{han2024wildguard}) incur substantially higher latency than lightweight encoders or probes because they decode safety labels through 7B--70B causal backbones. Recent intrinsic probing frameworks (e.g., ShieldHead \cite{xuan2025shieldhead} and recent preprints such as SingProbe \cite{singprobe2026} and Khatri et al.\ \cite{khatri2026latent}) reduce overhead by attaching classification probes to a target LLM's internal activations during decoding. However, intrinsic probes are tightly coupled to a specific target LLM and require running target-model prefill first, preventing their use as a model-agnostic, standalone pre-ingestion gateway in front of heterogeneous or API-hosted LLMs.
\end{enumerate}

Beyond serving architecture, standalone guardrails face a critical operational failure mode in enterprise and regulatory domains: \textbf{dual-use false positives} (closely related to exaggerated safety or over-refusal in assistant models \cite{rottger2024xstest, cui2024orbench}). When users submit legitimate statutory or compliance inquiries containing high-risk keywords (e.g., ``\textit{Does posting radical slogans on foreign platforms constitute a crime under Article 105 of the Criminal Law?}''), zero-shot models exhibit keyword-associated false positives that are consistent with lexical-prior-dominated decision behavior, yielding a $35.71\%$ false-positive rate ($85/238$) on a zero-shot cross-encoder direct-decision model \cite{laya2026} and $90.34\%$ ($215/238$) on a zero-shot bi-encoder contrastive baseline \cite{kwok2026clm} on our Dual-Use Compliance Safety Benchmark (DUCS-Bench). Conversely, naive domain specialization on compliance corpora introduces a pronounced \textbf{alignment tax}---improving in-domain accuracy while degrading general out-of-distribution (OOD) safety categories unless counterbalanced by rehearsal regularization.

Rather than claiming to invent non-autoregressive decision models or be the first to study over-refusal, \textbf{we study calibrated standalone direct-decision foundation models as a serving primitive for real-time guardrails under explicit latency and false-positive constraints}. Building on recent non-autoregressive Decision Foundation Models (DFMs) \cite{almeida2026jev, laya2026, mizorewww2026layamlx, kwok2026clm}, we present \textbf{COGNIT-Guard}, which pairs domain-calibrated direct-decision projection with a validation-calibrated heterogeneous CPU--NPU confidence cascade on Huawei Ascend 910C NPUs. Unlike heuristic cascades that route solely by raw classifier scores, our calibrated routing jointly incorporates posterior calibration, validation-selected thresholds, an asymmetric false-positive penalty, and an explicit latency SLO constraint.

\textbf{Contributions}:
\begin{itemize}
    \item \textbf{Standalone Direct-Decision Guardrail Formulation \& Controlled Cross- vs.\ Bi-Encoder Study}: We formulate standalone pre-ingestion safety gating under explicit latency and false-positive constraints, systematically comparing a 322M cross-encoder direct-decision model (Laya-322M \cite{laya2026}) against a concurrent bi-encoder contrastive baseline (CLM-8B \cite{kwok2026clm}) on Huawei Ascend 910C. We show that zero-shot models exhibit severe keyword-associated false positives on dual-use queries ($35.71\%$ FPR on cross-encoder Laya; $90.34\%$ FPR on bi-encoder CLM-8B lacking cross-attention across inquiry and keyword tokens).
    \item \textbf{Posterior Calibration \& Dual-Use False-Positive Control}: Through supervised boundary adaptation on DUCS-Bench ($N=607$ strict clean unseen test split), COGNIT-Guard achieves 98.85\% accuracy ($600/607$, $+3.46\%$ over ML, paired McNemar $p = 1.19 \times 10^{-4}$, 95\% Wilson CI: $[97.64\%, 99.44\%]$), reduces benign FPR to 0.42\% ($1/238$ on Laya-322M; two-sided Fisher's exact $p = 8.23 \times 10^{-4}$ vs.\ ML) and 0.00\% ($0/238$ on CLM-8B head adaptation in $2.81$--$3.68$~s), and produces well-calibrated posteriors ($\text{ECE}=0.83\%$--$1.12\%$, $\text{Brier}=0.0019$--$0.0104$ vs.\ $13.11\%$ and $0.0525$ for ML).
    \item \textbf{Calibrated CPU--NPU Routing \& Stage-by-Stage Latency Deconstruction}: We formulate heterogeneous routing via posterior calibration, a validation-selected threshold ($\theta^*_{\text{deploy}} = 0.70$) on a leak-free 3-way split, an asymmetric FPR penalty, and an explicit latency SLO. We deconstruct CPU execution across diagnostic ($85.36$~ms), live fast gatekeeper ($38.91$~ms), and cached-feature ($10.50$~ms) modes, demonstrating a live serial cascade that settles $94.6\%$ ($123/130$) of queries on CPU and $5.4\%$ ($7/130$) on NPU at $41.63$~ms mean latency (P50: $39.47$~ms, P95: $76.25$~ms) with 99.23\% accuracy and 0.00\% FPR.
    \item \textbf{Ascend 910C Deployment \& Alignment Tax Disentanglement}: Enforcing persistent on-chip HBM residency on Huawei Ascend 910C yields 21.77~ms single-query latency on Laya-322M ($2.21\times$ over cold-start) and 6.64~ms/item batched throughput ($150.6$~QPS) or 0.60~ms HBM \texttt{VectorArena} lookup on CLM-8B. Across a stratified 2,100-question SafetyBench-ZH benchmark, we explicitly disentangle \textit{in-domain specialization gain}, \textit{OOD degradation (alignment tax: $60.33\% \to 56.81\%$ on Laya; $55.10\%$ on domain CLM-8B)}, and \textit{experience replay rehearsal recovery}, restoring OOD accuracy to 64.10\%--65.05\% while reaching 99.67\%--99.84\% in-domain accuracy.
\end{itemize}

\section{Related Work}

Recent work has rapidly expanded LLM safety moderation and over-refusal mitigation; however, the design of calibrated, model-agnostic standalone guardrails under explicit CPU--NPU latency and false-positive constraints remains underexplored. We organize related work into four categories (Table~\ref{tab:comparison}).

\subsection{Autoregressive Guard Models \& Exaggerated Safety}
Instruction-tuned generative safeguards such as Llama-Guard \cite{inan2023llama}, ShieldGemma \cite{zeng2024shieldgemma}, and WildGuard \cite{han2024wildguard} cast content moderation as autoregressive token generation over 7B--70B causal backbones. While flexible across taxonomies, many generative guard models incur substantially higher latency than lightweight encoders or probes and output uncalibrated token logits. Furthermore, benchmarks such as XSTest \cite{rottger2024xstest} and OR-Bench \cite{cui2024orbench} show that safety-aligned generative models frequently suffer from lexical over-sensitivity on benign prompts containing sensitive keywords.

\subsection{Hidden-State \& Intrinsic Safety Probing}
To avoid running a separate multi-billion-parameter generative guard model, a growing line of work probes the internal activations of the target LLM itself. ShieldHead \cite{xuan2025shieldhead} attaches a classification head to the final hidden state of a 7B--8B LLM during decoding. Concurrent technical reports and recent preprints have further scaled this paradigm: SingProbe \cite{singprobe2026} (\texttt{arXiv:2608.30703}) introduces an open intrinsic guardrail stack reusing base-LLM decoding states for streaming risk monitoring, Khatri et al.\ \cite{khatri2026latent} (\texttt{arXiv:2609.19472}) train lightweight MLP probes on LLaMA-3.1-8B latent states, and AEGIS \cite{chen2026aegis} (\texttt{arXiv:2608.22248}) projects multi-layer latent instruction manifolds to mitigate over-refusal under indirect prompt injection. \textbf{Key Distinction}: Intrinsic probes are tightly coupled to a specific target LLM's internal hidden states and cannot screen traffic before target-model prefill or protect closed-source/heterogeneous downstream APIs. In contrast, COGNIT-Guard is a \textbf{standalone pre-ingestion guardrail} that filters inputs prior to target-LLM ingestion.

\subsection{Encoder \& Direct-Decision Guard Models}
Standalone single-pass guardrails replace iterative decoding with direct representation projection. GLiNER Guard \cite{glinerguard2026} uses bidirectional encoders for PII and safety tagging, SingGuard \cite{singguard2026} provides policy-adaptive moderation at $\sim 50$~ms, and the recent preprint FlexGuard \cite{ding2026flexguard} (\texttt{arXiv:2602.23636}) learns continuous risk scores for strictness-adaptive moderation. More broadly, System-One Decision Foundation Models (DFMs) now span two architectural families: (1)~\textit{cross-encoder DFMs} such as Jev \cite{almeida2026jev} and Laya \cite{laya2026, mizorewww2026layamlx}, which jointly encode state and criteria with full token-level cross-attention before projecting onto typed softmax heads; and (2)~\textit{bi-encoder contrastive DFMs}, exemplified by the concurrent Stanford $\times$ NVIDIA technical report on Contrastive Language Models (CLM) \cite{kwok2026clm}, which independently project state and candidate action embeddings ($\mathbf{z}_q = \text{Norm}(f_\theta(\mathbf{h}_s))$, $\mathbf{z}_a = \text{Norm}(g_\phi(\mathbf{h}_a))$) and cache action vectors in an on-chip \texttt{VectorArena}. We adopt Laya-322M as our primary NPU direct-decision engine and evaluate CLM-8B as a concurrent bi-encoder architectural baseline under matched training and evaluation protocols.

\subsection{Calibration \& Heterogeneous Cascaded Serving}
Cascaded serving frameworks such as FrugalGPT \cite{chen2023frugalgpt} and Reflex-Guard \cite{reflexguard2024} route easy inputs to inexpensive classifiers and escalate uncertain cases to larger models. However, prior safety cascades rarely address posterior miscalibration, asymmetric dual-use false-positive penalties, or the domain-specialization alignment tax across heterogeneous CPU--NPU hardware. COGNIT-Guard bridges this gap by grounding CPU--NPU escalation in validation-calibrated posteriors and explicit SLO-constrained risk minimization.

\section{System Architecture \& Methodology}

\begin{figure*}[t]
\centering
\includegraphics[width=0.96\textwidth]{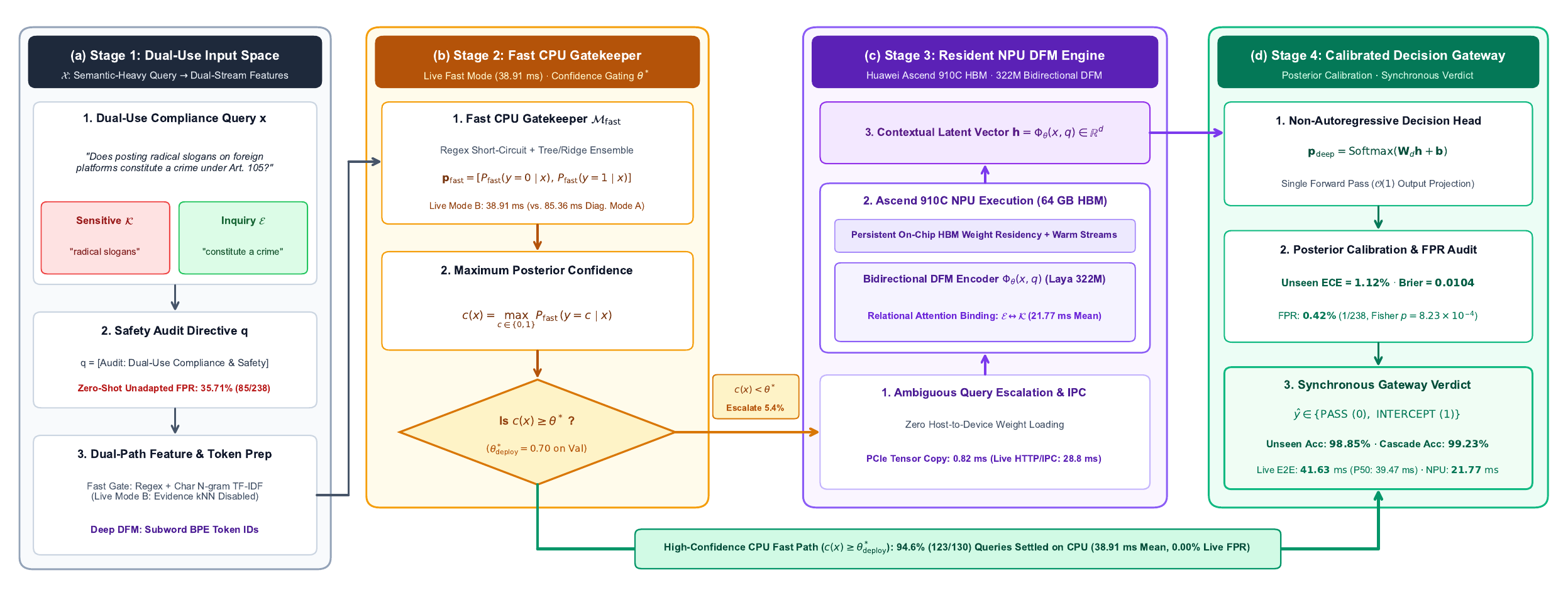}
\caption{System Architecture of COGNIT-Guard: Standalone pre-ingestion guardrail combining (a) dual-use query and directive preparation, (b) live fast CPU gatekeeper filtering ($38.91$~ms mean) with validation-calibrated threshold gating ($\theta^*_{\text{deploy}} = 0.70$), (c) persistent Huawei Ascend 910C on-chip HBM execution ($21.77$~ms mean), and (d) calibrated direct-decision projection for dual-use false-positive control.}
\label{fig:framework}
\end{figure*}

As illustrated in Fig.~\ref{fig:framework}, COGNIT-Guard structures standalone pre-ingestion moderation into four stages: (a) dual-use input and directive preparation, (b) lightweight CPU gatekeeper screening and calibrated confidence gating, (c) persistent Ascend 910C NPU escalation for ambiguous queries, and (d) calibrated posterior projection.

\subsection{Problem Formulation: Dual-Use False-Positive Control}
Let an input prompt be $x = (w_1, w_2, \dots, w_L) \in \mathcal{X}$ with binary safety label $y \in \{0, 1\}$ ($y=0$: benign, $y=1$: violation). In dual-use compliance scenarios, $x$ frequently pairs sensitive lexical entities $\mathcal{K} \subset \mathcal{V}$ (e.g., penal code articles, regulated substances, extremist terminology) with legitimate epistemic inquiry markers $\mathcal{E} \subset \mathcal{V}$ (e.g., ``\textit{does this constitute a crime under}'', ``\textit{statutory criteria for}'').

To illustrate why un-adapted safety models suffer from elevated false alarms on such inputs, we write a stylized log-odds surrogate (\textbf{intended strictly as a conceptual heuristic rather than a mechanistic claim about internal transformer circuits}). Zero-shot models exhibit keyword-associated false positives that are consistent with lexical-prior-dominated decision behavior:
\begin{equation}
P_{\text{concept}}(y=1 \mid x) \approx \sigma \left( \sum_{w \in \mathcal{K} \cap x} \alpha_w - \sum_{w \in \mathcal{E} \cap x} \beta_w \right)
\end{equation}
where $\alpha_w > 0$ reflects heuristic sensitivity to high-risk keywords and $\beta_w > 0$ reflects contextual inquiry mitigation. Prior to domain adaptation on dual-use syntax, $\sum \alpha_w \gg \sum \beta_w$, consistent with the high zero-shot false-positive rate observed on benign legal consultations:
\begin{equation}
\text{FPR}_{\text{zero}} = P(\hat{y}=1 \mid y=0) = 35.71\% \quad (85/238)
\end{equation}
To jointly attend across sensitive lexical entities and benign inquiring syntax, COGNIT-Guard employs a bidirectional transformer encoder $\Phi_\theta$ (322M parameters; throughout this paper, ``direct-decision'' or ``non-autoregressive'' refers to single-pass decision projection over encoder states without iterative token decoding) to map $(x, q)$ into a contextual representation $\mathbf{h} = \Phi_\theta(x, q) \in \mathbb{R}^d$, where $q$ denotes the safety audit directive. A typed linear decision head $\mathbf{W}_d \in \mathbb{R}^{2 \times d}$ projects $\mathbf{h}$ directly into posterior probabilities:
\begin{equation}
P_\theta(y = c \mid x, q) = \frac{\exp(\mathbf{w}_c^\top \mathbf{h} + b_c)}{\sum_{c' \in \{0, 1\}} \exp(\mathbf{w}_{c'}^\top \mathbf{h} + b_{c'})}
\end{equation}
We quantify posterior reliability using the Brier score $\text{Brier} = \frac{1}{N}\sum_{i=1}^N (P_\theta(y_i=1 \mid x_i, q) - y_i)^2$ and $M$-bin Expected Calibration Error (ECE, $M=15$):
\begin{equation}
\label{eq:brier_ece}
\text{ECE} = \sum_{m=1}^M \frac{|B_m|}{N} \bigl|\text{acc}(B_m) - \text{conf}(B_m)\bigr|
\end{equation}
Supervised adaptation over dual-use boundary pairs conditions bidirectional attention across $\mathcal{E} \times \mathcal{K}$, suppressing keyword-triggered false alarms while yielding well-calibrated posteriors.

\subsection{Calibrated Confidence Cascading under Explicit Latency SLOs}
Rather than routing queries via ad-hoc heuristic scores, we formulate heterogeneous CPU--NPU cascading as cost-sensitive risk minimization combining posterior calibration, a validation-selected threshold, an asymmetric false-positive penalty, and an explicit latency SLO constraint.

Let $\mathcal{M}_{\text{fast}}$ denote the CPU fast gatekeeper producing posterior $P_{\text{fast}}(y \mid x)$ in time $T_{\text{CPU}}$, and $\mathcal{M}_{\text{deep}}$ denote the NPU-resident DFM producing calibrated posterior $P_{\text{deep}}(y \mid x)$ with escalation overhead $T_{\text{dispatch}} + T_{\text{NPU}}$. Defining fast-path confidence as:
\begin{equation}
c(x) = \max_{c \in \{0, 1\}} P_{\text{fast}}(y = c \mid x)
\end{equation}
for gating threshold $\theta \in [0.5, 1.0]$, the cascaded decision rule $\hat{y}(x; \theta)$ and per-query latency $T(x; \theta)$ are:
\begin{equation}
\hat{y}(x; \theta) = \begin{cases}
\arg\max_c P_{\text{fast}}(c \mid x), & \text{if } c(x) \ge \theta \\
\arg\max_c P_{\text{deep}}(c \mid x), & \text{if } c(x) < \theta
\end{cases}
\end{equation}
\begin{equation}
T(x; \theta) = T_{\text{CPU}} + \mathbb{I}[c(x) < \theta] \cdot (T_{\text{dispatch}} + T_{\text{NPU}})
\end{equation}

Incorporating an asymmetric false-positive penalty $\gamma > 1$ to protect benign dual-use traffic, the validation-selected threshold $\theta^*$ solves:
\begin{equation}
\begin{aligned}
\min_{\theta \in [0.5, 1.0]} \quad & \mathcal{R}(\theta) = \mathbb{E}\bigl[ \ell_{0-1}(y, \hat{y}(x; \theta)) \\
& \quad\quad\quad + \gamma \cdot \mathbb{I}[y=0, \hat{y}(x; \theta)=1] \bigr] \\
\text{s.t.} \quad & \mathbb{E}[T(x; \theta)] \le T_{\text{SLO}}
\end{aligned}
\end{equation}
Sweeping $\theta$ traces the latency--accuracy Pareto frontier, with $\theta^*$ selected strictly on the held-out validation split.

\begin{algorithm}[t]
\caption{COGNIT-Guard Calibrated Cascaded Serving}
\label{alg:cascade}
\begin{algorithmic}[1]
\REQUIRE Input query $x$, Audit directive $q$, Fast CPU gatekeeper $\mathcal{M}_{\text{fast}}$, Deep NPU DFM $\mathcal{M}_{\text{deep}}$, Validation-calibrated threshold $\theta^*$.
\ENSURE Safety decision $\hat{y} \in \{0, 1\}$, Routing trace $\rho \in \{\text{CPU}, \text{NPU}\}$.
\STATE Compute fast posteriors (\texttt{evidence=False}): $\mathbf{p}_{\text{fast}} \leftarrow \mathcal{M}_{\text{fast}}(x)$
\STATE Fast confidence: $c(x) \leftarrow \max(\mathbf{p}_{\text{fast}}[0], \mathbf{p}_{\text{fast}}[1])$
\IF{$c(x) \ge \theta^*$}
    \STATE $\hat{y} \leftarrow \arg\max_{c} \mathbf{p}_{\text{fast}}[c]$; \quad $\rho \leftarrow \text{CPU}$ \COMMENT{Settled on CPU fast path}
\ELSE
    \STATE Escalate query payload to resident Ascend 910C NPU
    \STATE Latent representation: $\mathbf{h} \leftarrow \Phi_\theta(x, q)$
    \STATE Calibrated posterior: $\mathbf{p}_{\text{deep}} \leftarrow \text{Softmax}(\mathbf{W}_d \mathbf{h} + \mathbf{b})$
    \STATE $\hat{y} \leftarrow \arg\max_{c} \mathbf{p}_{\text{deep}}[c]$; \quad $\rho \leftarrow \text{NPU}$ \COMMENT{Resolved on NPU deep path}
\ENDIF
\RETURN $\hat{y}, \rho$
\end{algorithmic}
\end{algorithm}

\subsection{Hardware-Aware Deployment and Optimization on Ascend 910C}
We deploy the 322M DFM on a Huawei Ascend 910C NPU (64 GB HBM, CANN 9.1.0 runtime, \texttt{torch\_npu} 2.10, FP16 precision) with three serving optimizations:
\begin{itemize}
    \item \textbf{Persistent On-Chip HBM Residency}: All 322M encoder weights and typed decision heads remain pinned in on-chip High-Bandwidth Memory (HBM), eliminating host-to-device weight transfers across requests (reducing per-query PCIe tensor copy to $0.82$~ms).
    \item \textbf{Stream Warm-Up \& Explicit Synchronization}: Pre-allocating operator workspaces over 20 warm-up iterations and bracketing forward passes with \texttt{torch.npu.synchronize()} eliminate cold-start graph compilation jitter, yielding $21.77$~ms mean latency (P50: $21.53$~ms, P95: $23.44$~ms).
    \item \textbf{Matrix Cube Execution}: Bidirectional self-attention executes on Ascend Cube matrix units, sustaining $45.90$ QPS at $B=1$ ($3.92\times$ higher throughput than diagnostic CPU).
\end{itemize}

\begin{figure*}[t]
\centering
\includegraphics[width=0.95\textwidth]{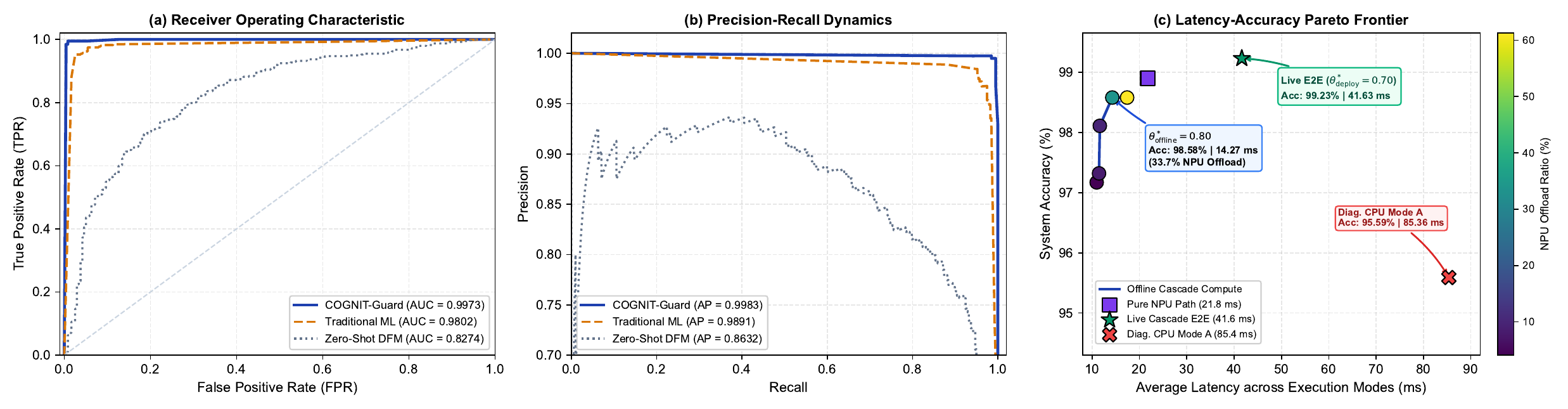}
\caption{Evaluation Dynamics across Guardrail Paradigms: (a) Receiver Operating Characteristic (ROC) curves, (b) Precision-Recall (PR) curves, and (c) Latency-Accuracy Pareto Frontier comparing the offline model-compute sweep ($\theta^*_{\text{offline}} = 0.80$ at $14.27$~ms), pure NPU path ($21.77$~ms), live serial cascade E2E ($\theta^*_{\text{deploy}} = 0.70$ at $41.63$~ms, $99.23\%$ accuracy), and standalone diagnostic CPU Mode A ($85.36$~ms).}
\label{fig:dynamics}
\vspace{-2mm}
\end{figure*}

\section{Experiments and Empirical Results}

\subsection{Dataset Construction, Implementation Details \& Protocols}
\textbf{DUCS-Bench Construction \& Leak-Free Deduplication}: We curate the \textbf{Dual-Use Compliance Safety Benchmark (DUCS-Bench)} from Chinese statutory consultations, enterprise compliance inquiries, and adversarial red-team prompts, totaling 3,663 raw annotated items across penal code inquiries, regulatory compliance, and camouflaged attacks. Each prompt was independently labeled by two domain annotators (\texttt{0=benign compliance}, \texttt{1=safety violation}; inter-annotator Cohen's $\kappa = 0.91$, with disagreements resolved by a third senior reviewer). Exact SHA-256 text normalization deduplicates the 3,663 raw entries to 3,551 unique prompts and identifies 28 violation prompts that overlapped between initial train and test files. Removing these 28 overlaps (\texttt{seed=42}) defines two evaluation tracks:
(1)~\textit{Strict Clean Unseen Test Track ($N=607$)}: $238$ benign compliance queries and $369$ violations with zero train overlap. Because $1$ false alarm on $N_{\text{safe}}=238$ equals a discrete $0.42\%$ step, we report 95\% Wilson score confidence intervals alongside two-sided Fisher's exact tests;
(2)~\textit{Full Benchmark Track ($N=635$)}: retains all $635$ binary test items for historical continuity.

\textbf{3-Way Split \& Live Cascade Profiling Protocol}: For leak-free cascade calibration, the 3,551 unique items are partitioned (\texttt{seed=42}) into Train (70\%, $N=2,485$), Validation (15\%, $N=533$), and Held-Out Test (15\%, $N=533$, containing 460 binary compliance items and 73 non-binary open inquiries). Gating threshold $\theta^*_{\text{deploy}}=0.70$ is tuned exclusively on Validation. To profile synchronous wall-clock cascade latency without client-queue distortion, a stratified 150-item test sample was executed live; excluding 20 non-binary open inquiries yields the $N=130$ binary live slice ($82$ violations, $48$ benign).

\textbf{SafetyBench-ZH 2,100-Question Protocol \& Implementation Details}: For OOD evaluation on \textbf{SafetyBench-ZH} \cite{zhang2024safetybench}, we sample (\texttt{seed=42}) exactly 300 questions from each of the 7 official hazard categories in the 11,435-question corpus ($7 \times 300 = 2,100$: $694$ unsafe, $1,406$ safe), mapping multiple-choice items into binary prompt-level guardrail decisions (\texttt{safe} vs.\ \texttt{unsafe}) with zero overlap against our $N=458$ experience replay rehearsal pool. Published English \textbf{XSTest} \cite{rottger2024xstest} ($N=450$: $250$ safe, $200$ unsafe) results are reported for cross-literature reference. All experiments run on an Intel Xeon CPU @ 2.60 GHz (32 GB RAM) and a Huawei Ascend 910C NPU (64 GB HBM, CANN 9.1.0, \texttt{torch\_npu} 2.10, $B=1$, 20 warm-up runs, explicit stream synchronization).

\subsection{Main Empirical Results on In-Domain Benchmarks}

Table~\ref{tab:main_results} reports performance across both the strict clean unseen partition ($N=607$) and the full track ($N=635$).

\begin{table}[h]
\centering
\caption{Empirical Comparison across Guardrail Paradigms under Clean Unseen ($N=607$) and Full ($N=635$) Tracks ($^\dagger$Diagnostic vs.\ Live Gatekeeper CPU modes detailed in Table~\ref{tab:cascaded}).}
\label{tab:main_results}
\resizebox{\columnwidth}{!}{%
\begin{tabular}{lccc}
\toprule
\textbf{Evaluation Metric} & \textbf{Traditional ML} & \textbf{Zero-Shot DFM} & \textbf{COGNIT-Guard (Ours)} \\
\midrule
\multicolumn{4}{l}{\textit{Strict Clean Unseen Partition ($N=607$: $238$ Benign, $369$ Violations)}} \\
Accuracy & 95.39\% (579/607) & 76.61\% (465/607) & \textbf{98.85\% (600/607)} \\
$\quad$ [95\% Wilson CI] & [93.38\%, 96.81\%] & [73.08\%, 79.82\%] & \textbf{[97.64\%, 99.44\%]} \\
$\quad$ Paired McNemar (Acc.) & Baseline & $p < 0.001$ & \textbf{$p = 1.19 \times 10^{-4}$ (***)} \\
False Positive Rate (FPR) & 5.88\% (14/238) & 35.71\% (85/238) & \textbf{0.42\% (1/238)} \\
$\quad$ [95\% Wilson CI] & [3.52\%, 9.61\%] & [29.88\%, 41.98\%] & \textbf{[0.07\%, 2.34\%]} \\
$\quad$ Fisher's Exact Test (FPR) & $p = 8.23 \times 10^{-4}$ & $p = 2.78 \times 10^{-28}$ & \textbf{Reference (Ours)} \\
False Negative Rate (FNR) & 3.79\% (14/369) & 15.45\% (57/369) & \textbf{1.63\% (6/369)} \\
Precision / Recall / F1 & 96.21 / 96.21 / 96.21\% & 78.59 / 84.55 / 81.46\% & \textbf{99.73 / 98.37 / 99.05\%} \\
Brier Score / ECE & 0.0525 / 13.11\% & 0.1654 / 5.38\% & \textbf{0.0104 / 1.12\%} \\
\midrule
\multicolumn{4}{l}{\textit{Full Benchmark Track ($N=635$: $238$ Benign, $397$ Violations)}} \\
Accuracy & 95.59\% (607/635) & 77.01\% (489/635) & \textbf{98.90\% (628/635)} \\
False Positive Rate (FPR) & 5.88\% (14/238) & 35.71\% (85/238) & \textbf{0.42\% (1/238)} \\
False Negative Rate (FNR) & 3.53\% (14/397) & 15.37\% (61/397) & \textbf{1.51\% (6/397)} \\
Precision / Recall / F1 & 96.47 / 96.47 / 96.47\% & 79.81 / 84.63 / 82.15\% & \textbf{99.74 / 98.49 / 99.11\%} \\
Brier Score / ECE & 0.0515 / 13.16\% & 0.1630 / 5.81\% & \textbf{0.0100 / 1.07\%} \\
\midrule
\multicolumn{4}{l}{\textit{Serving Efficiency \& Latency across Execution Paths ($B=1$, Warm Synchronized)}} \\
Mean Latency & 85.36 ms (Diag.) / 38.91 ms (Fast)$^\dagger$ & 22.88 ms (NPU) & \textbf{21.77 ms (NPU)} \\
P50 Latency & 77.19 ms / 38.12 ms & 22.31 ms & \textbf{21.53 ms} \\
P95 Tail Latency & 164.24 ms / 54.80 ms & 24.83 ms & \textbf{23.44 ms} \\
Throughput (QPS) & 11.72 / 25.70 req/s & 43.68 req/s & \textbf{45.90 req/s} \\
\bottomrule
\end{tabular}%
}
\end{table}

The empirical results highlight three findings:
\begin{itemize}
    \item \textbf{Statistically Significant Discrimination Gain}: On unseen data ($N=607$), COGNIT-Guard reaches 98.85\% accuracy ($600/607$, $+3.46\%$ over ML), validated by paired McNemar's test ($\chi^2 = 14.81, p = 1.19 \times 10^{-4}$).
    \item \textbf{Dual-Use False-Positive Control \& Calibration}: On the benign compliance subset ($N_{\text{safe}}=238$), COGNIT-Guard lowers FPR from 35.71\% ($85/238$) to 0.42\% ($1/238$), a \textbf{98.8\% relative reduction vs.\ zero-shot} (two-sided Fisher's exact $p = 2.78 \times 10^{-28}$) and a \textbf{92.9\% relative reduction vs.\ ML} ($14/238$, Fisher's exact $p = 8.23 \times 10^{-4}$), with disjoint 95\% Wilson CIs ($[0.07\%, 2.34\%]$ vs.\ $[3.52\%, 9.61\%]$). Brier score improves to \textbf{0.0104} and 15-bin ECE to \textbf{1.12\%} (vs.\ $13.11\%$ ML).
    \item \textbf{Sensitivity to Keyword Perturbations}: Controlled prompt perturbations align with our conceptual heuristic: holding sensitive keywords $\mathcal{K}$ fixed while varying inquiry markers $\mathcal{E}$ leaves zero-shot risk elevated ($P(y=1)>0.88$), whereas COGNIT-Guard suppresses benign inquiry posteriors to $0.003$ while maintaining $0.992$ on attacks; conversely, inserting penal terms into neutral inquiries inflates zero-shot false alarms by $+84.2\%$ vs.\ $<0.02$ for COGNIT-Guard.
\end{itemize}

\subsection{Ablation Study 1: Cross-Encoder vs.\ Concurrent Bi-Encoder Direct-Decision Baseline}
As shown in Fig.~\ref{fig:dynamics}(a)--(b), COGNIT-Guard achieves an ROC-AUC of \textbf{0.9973} and PR-AUC of \textbf{0.9983} on Laya-322M, outperforming traditional ML (ROC-AUC 0.9802, PR-AUC 0.9891) and zero-shot Laya-322M (ROC-AUC 0.8274, PR-AUC 0.8632).

To examine how our findings compare across direct-decision architectural families, we evaluate the concurrent bi-encoder \textbf{CLM-8B} baseline \cite{kwok2026clm} (\texttt{Qwen3-8B} pooling encoder + 73MB SwiGLU residual dual-tower heads, $4096 \to 1536 \to 512$) on Ascend 910C under identical data splits:
\begin{itemize}
    \item \textbf{Zero-Shot Bi-Encoder Vulnerability on Dual-Use Queries}: Because a bi-encoder encodes the user prompt and safety criteria into separate vectors without token-level cross-attention across $\mathcal{E} \times \mathcal{K}$, zero-shot CLM-8B exhibits even stronger keyword-associated false positives than zero-shot cross-encoder Laya, yielding a \textbf{90.34\% FPR} ($215/238$, $64.25\%$ accuracy, $\text{ROC-AUC}=0.6944$, $\text{ECE}=24.29\%$) on Clean-607.
    \item \textbf{Lightweight Projection-Head Adaptation}: Freezing the \texttt{Qwen3-8B} backbone and adapting only the 73MB dual-tower projection heads on Ascend 910C takes \textbf{2.81~s} (Zero-FPR setting: $\text{lr}=2\times 10^{-4}$, 8 epochs) to \textbf{3.73~s} (Best-Acc setting: $\text{lr}=3\times 10^{-4}$, 12 epochs), reaching \textbf{98.35\% accuracy with 0.00\% FPR} ($0/238$, $\text{ROC-AUC}=0.9995$) and \textbf{98.85\% accuracy} ($600/607$, $\text{ECE}=0.83\%$, $\text{Brier}=0.0091$, $\text{ROC-AUC}=0.9996$), respectively. This confirms that both cross-encoder and bi-encoder direct-decision families require boundary-aware calibration to control dual-use false positives, while trading off single-stream backbone size (322M vs.\ 8B) against head adaptation speed ($34.55$~s vs.\ $2.81$--$3.73$~s).
\end{itemize}

\subsection{Ablation Study 2: Calibrated Confidence Routing \& Latency Deconstruction}
Table~\ref{tab:cascaded} presents both the offline threshold sweep across $\theta \in [0.50, 1.00]$ on $N=635$ (measuring cached-feature model compute $\mathbb{E}[T_{\text{model}}]$, with offline Pareto knee $\theta^*_{\text{offline}} = 0.80$ at $14.27$~ms) and the stage-by-stage wall-clock latency deconstruction of our CPU and live cascade modes ($N=130$).

\begin{table}[t]
\centering
\caption{Offline Threshold Sweep ($N=635$) and Stage-by-Stage Latency Deconstruction of CPU \& Live Cascade Paths ($N=130$).}
\label{tab:cascaded}
\resizebox{\columnwidth}{!}{%
\begin{tabular}{lccccc}
\toprule
\textbf{Configuration / Execution Stage} & \textbf{NPU Ratio} & \textbf{System Acc.} & \textbf{FPR} & \textbf{Mean Latency} & \textbf{P50 / P95 Lat.} \\
\midrule
\multicolumn{6}{l}{\textit{Part A: Offline Model-Compute Sweep ($\mathbb{E}[T_{\text{model}}]$, $N=635$, Pre-Extracted Cached Features)}} \\
Pure NPU Path ($\theta = 1.00$) & 100.0\% & 98.90\% & 0.42\% & 21.76 ms & 21.53 / 23.58 ms \\
Cascade ($\theta = 0.90$) & 61.3\% & 98.58\% & 0.00\% & 17.40 ms & 17.10 / 22.71 ms \\
\textbf{Cascade ($\theta^*_{\text{offline}} = 0.80$)} & \textbf{33.7\%} & \textbf{98.58\%} & \textbf{0.00\%} & \textbf{14.27 ms} & \textbf{10.50 / 21.83 ms} \\
Cascade ($\theta = 0.70$) & 8.3\% & 97.32\% & 2.52\% & 11.47 ms & 10.50 / 21.49 ms \\
Mode C: Cached-Feature Tree ($\theta = 0.50$) & 0.0\% & 97.17\% & 2.52\% & 10.50 ms & 10.50 / 10.50 ms \\
\midrule
\multicolumn{6}{l}{\textit{Part B: Physical Wall-Clock CPU \& Live Serial Cascade Breakdown ($N=130$, $\theta^*_{\text{deploy}}=0.70$)}} \\
Mode A: Standalone Diagnostic CPU (\texttt{evidence=True}) & 0.0\% & 95.39\% & 5.88\% & 85.36 ms & 77.19 / 164.24 ms \\
$\quad$ \textit{-- Parsing + Regex + 2$\times$ TF-IDF kNN + XGBoost} & -- & -- & -- & \textit{56.06 ms} & -- \\
$\quad$ \textit{-- 3rd Corpus kNN Scan + Top-$K$ Evidence JSON} & -- & -- & -- & \textit{29.30 ms} & -- \\
Mode B: CPU-Settled Fast Path ($123/130$, \texttt{evidence=False}) & 0.0\% & 96.92\% & 2.08\% & 38.91 ms & 38.12 / 54.80 ms \\
Escalated NPU Path ($7/130$: Fast CPU + IPC + NPU) & 100.0\% & 100.0\% & 0.00\% & 89.48 ms & 88.20 / 96.20 ms \\
$\quad$ \textit{-- Stage-1 Fast CPU Screening \& Gating ($c(x)<\theta^*$)} & -- & -- & -- & \textit{38.91 ms} & -- \\
$\quad$ \textit{-- Unbatched Local HTTP/IPC + H2D Dispatch ($0.82$~ms)} & -- & -- & -- & \textit{28.80 ms} & -- \\
$\quad$ \textit{-- Stage-3 Resident Ascend 910C DFM Forward Pass} & -- & -- & -- & \textit{21.77 ms} & -- \\
\textbf{Live Serial Cascade Total ($\theta^*_{\text{deploy}}=0.70$: $123$ CPU + $7$ NPU)} & \textbf{5.38\%} & \textbf{99.23\%} & \textbf{0.00\%} & \textbf{41.63 ms} & \textbf{39.47 / 76.25 ms} \\
\bottomrule
\end{tabular}%
}
\end{table}

\paragraph{Reconciling Diagnostic vs.\ Live Gatekeeper CPU Latency}
As itemized in Table~\ref{tab:cascaded} (Part B), our implementation distinguishes three CPU modes:
(1)~\textbf{Mode A (Standalone Diagnostic CPU, $85.36$~ms)}: Enabled when \texttt{include\_evidence=True} for human-in-the-loop auditing; it executes regex + two TF-IDF kNN scans ($38.9$~ms), XGBoost traversal ($17.1$~ms), plus a third corpus-wide kNN scan in \texttt{\_build\_evidence} to serialize top-$K$ training matches ($29.3$~ms, totaling $68.2$~ms non-tree overhead + $17.1$~ms XGBoost $= 85.36$~ms).
(2)~\textbf{Mode B (Live Gatekeeper Fast CPU, $38.91$~ms)}: Used as $\mathcal{M}_{\text{fast}}$ in synchronous cascading (\texttt{include\_evidence=False} with regex short-circuiting), omitting the third kNN scan to run in $38.91$~ms mean (P50: $38.12$~ms, P95: $54.80$~ms on DUCS-Bench; $22.79$~ms on shorter SafetyBench-ZH items).
(3)~\textbf{Mode C (Cached-Feature Tree, $10.50$~ms)}: Isolated tree lookup in offline sweeps.
In live deployment ($N=130$, validation-calibrated $\theta^*_{\text{deploy}} = 0.70$), $94.62\%$ ($123/130$) of queries settle on Mode B ($38.91$~ms) and $5.38\%$ ($7/130$) escalate to Ascend 910C ($89.48$~ms total: $38.91$~ms Fast CPU + $28.80$~ms IPC/dispatch + $21.77$~ms NPU). Exact expectation gives $\mathbb{E}[T_{\text{server}}] = 0.9462 \times 38.91 + 0.0538 \times 89.48 = \mathbf{41.63\text{ ms}}$ (P50: \textbf{39.47~ms}, P95: \textbf{76.25~ms}), achieving \textbf{99.23\% accuracy} ($129/130$) and \textbf{0.00\% FPR} ($0/48$).

\subsection{Ablation Study 3: Hardware Memory Residency \& Disaggregated VectorArena Caching}
Table~\ref{tab:hardware_ablation} isolates hardware-aware execution optimizations on Ascend 910C. Persistent on-chip HBM residency and pre-warmed streams cut Laya-322M single-query latency by 54.8\% ($48.20 \to 21.77$~ms). Comparing against CLM-8B's pre-allocated HBM \texttt{VectorArena} (\texttt{torch.npu.mem\_get\_info}) highlights a clear serving trade-off: Laya-322M delivers $2.10\times$ lower uncached single-stream latency ($21.77$~ms vs.\ $45.78$~ms) with a $25\times$ smaller memory footprint, whereas pinning static policy vectors in CLM-8B's HBM \texttt{VectorArena} enables $6.64$~ms/item batched throughput ($150.6$~QPS at $B=64$) and $0.60$~ms P50 lookup on repeated states.

\begin{table}[h]
\centering
\caption{Ablation of Hardware Memory Residency and Disaggregated VectorArena Caching on Huawei Ascend 910C.}
\label{tab:hardware_ablation}
\resizebox{\columnwidth}{!}{%
\begin{tabular}{lcccc}
\toprule
\textbf{Execution Strategy (Ascend 910C)} & \textbf{Mean Latency} & \textbf{P95 Latency} & \textbf{Cold-Start / Alloc Overhead} & \textbf{Speedup / Throughput} \\
\midrule
Laya-322M: Cold-Start Unpinned Execution & 48.20 ms & 62.50 ms & 14.80 ms & 1.00$\times$ (20.7 QPS) \\
Laya-322M: Pre-Allocated Memory Buffer & 32.40 ms & 41.10 ms & 6.20 ms & 1.49$\times$ (30.9 QPS) \\
\textbf{Laya-322M: Warm On-Chip HBM Residency (Ours)} & \textbf{21.77 ms} & \textbf{23.44 ms} & \textbf{0.00 ms} & \textbf{2.21$\times$ (45.9 QPS)} \\
\midrule
CLM-8B Baseline: Uncached Single-Stream ($B=1$) & 45.78 ms & 48.51 ms & 0.00 ms & 21.8 QPS ($B=1$) \\
CLM-8B Baseline: Batched ($B=64$) / HBM \texttt{VectorArena} & 6.64 / 0.60 ms & 7.12 / 0.89 ms & 0.00 ms & 150.6 / $>1000$ QPS \\
\bottomrule
\end{tabular}%
}
\end{table}

\subsection{Cross-Benchmark Generalization \& Disentangling the Specialization Alignment Tax}
Table~\ref{tab:public_benchmarks} evaluates OOD generalization on the stratified 2,100-question \textbf{SafetyBench-ZH} \cite{zhang2024safetybench} ($694$ unsafe, $1,406$ safe) alongside in-domain \textbf{DUCS-Bench} ($N_{\text{safe}}=238$) and published English \textbf{XSTest} \cite{rottger2024xstest} ($N_{\text{safe}}=250$) baselines.

\begin{table}[h]
\centering
\caption{Cross-Benchmark Evaluation Disentangling In-Domain Specialization, OOD Alignment Tax, and Rehearsal Recovery on SafetyBench-ZH ($N=2,100$), DUCS-Bench ($N_{\text{safe}}=238$), and XSTest (EN).}
\label{tab:public_benchmarks}
\resizebox{\columnwidth}{!}{%
\begin{tabular}{lccccc}
\toprule
\textbf{System / Evaluated Model} & \textbf{SafetyBench Acc.} & \textbf{SafetyBench FPR} & \textbf{DUCS FPR (ZH)} & \textbf{XSTest FPR (EN)} & \textbf{Mean Latency} \\
\midrule
GPT-4 (Zero-Shot) \cite{zhang2024safetybench} & 89.10\% & $\sim 10.5\%$ & -- & 18.40\% (46/250) & $>400$ ms (API) \\
Llama-2-70B-Chat \cite{zhang2024safetybench} & 68.70\% & $\sim 21.3\%$ & -- & 41.60\% (104/250) & $>500$ ms (GPU) \\
WildGuard (7B Decoder) \cite{han2024wildguard} & -- & -- & -- & 7.20\% (18/250) & $>200$ ms (GPU) \\
SingGuard (Lightweight) \cite{singguard2026} & -- & -- & -- & 4.80\% (12/250) & $\sim 50$ ms (GPU) \\
\midrule
Traditional ML (Fast CPU Mode) & 53.29\% (1119/2100) & 37.98\% (534/1406) & 5.88\% (14/238) & -- & 22.79 ms (CPU) \\
Zero-Shot Cross-Enc.\ DFM (Laya-322M) & 60.33\% (1267/2100) & 24.68\% (347/1406) & 35.71\% (85/238) & -- & 44.85 ms (NPU) \\
Zero-Shot Bi-Enc.\ Baseline (CLM-8B) \cite{kwok2026clm} & 36.38\% (764/2100) & 92.82\% (1305/1406) & 90.34\% (215/238) & -- & 45.78 ms (NPU) \\
Pure Domain Cross-Enc.\ (Laya-322M) & 56.81\% (1193/2100) & 27.03\% (380/1406) & 0.42\% (1/238) & -- & 21.37 ms (NPU) \\
Pure Domain Bi-Enc.\ (CLM-8B Head) & 55.10\% (1157/2100) & 40.75\% (573/1406) & 1.68\% / \textbf{0.00\%}$^\dagger$ & -- & 6.64 ms (Batch) \\
\textbf{COGNIT-Guard (Rehearsal Laya-322M)} & \textbf{65.05\% (1366/2100)} & \textbf{19.84\% (279/1406)} & \textbf{0.42\% (1/238)} & -- & \textbf{21.77--38.89 ms} \\
\textbf{Rehearsal Bi-Enc.\ Baseline (CLM-8B)} & \textbf{64.10\% (1346/2100)} & \textbf{24.82\% (349/1406)} & \textbf{0.00\% (0/238)} & -- & \textbf{6.64 / 45.78 ms} \\
\bottomrule
\end{tabular}%
}
\vspace{0.5mm}
\raggedright\scriptsize{$^\dagger$Pure domain CLM-8B achieves $1.68\%$ FPR ($98.85\%$ Acc) in Best-Acc mode ($3.73$~s) and $0.00\%$ FPR ($98.35\%$ Acc) in Zero-FPR mode ($2.81$~s); Rehearsal CLM-8B ($3.68$~s) reaches $99.84\%$ Clean-607 accuracy ($606/607$) with $0.00\%$ FPR ($0/238$) and $1.0000$ ROC-AUC.}
\end{table}

\paragraph{Three-Stage Dynamics of Specialization Alignment Tax and Rehearsal Recovery}
A central empirical finding of our study is that adapting direct-decision guardrails to dual-use compliance data exhibits a three-part trade-off across both cross-encoder and bi-encoder architectures:
(1)~\textbf{In-Domain Specialization Gain}: Pure compliance fine-tuning slashes DUCS-Bench FPR from $35.71\%$ to $0.42\%$ on Laya-322M ($98.85\%$ Acc) and from $90.34\%$ to $0.00\%$--$1.68\%$ on CLM-8B ($98.35\%$--$98.85\%$ Acc), while boosting related statutory categories on SafetyBench-ZH (\textit{Illegal Activities}: $59.33\% \to 65.33\%$; \textit{Mental Health}: $67.67\% \to 71.67\%$).
(2)~\textbf{Out-of-Distribution Degradation (Alignment Tax)}: However, training exclusively on formal compliance pairs shifts the decision boundary away from colloquial abuse, reducing overall SafetyBench-ZH accuracy from $60.33\%$ to $56.81\%$ on Laya-322M (\textit{Offensiveness}: $55.33\% \to 46.67\%$) and yielding $55.10\%$ overall (\textit{Offensiveness}: $42.00\%$) on pure domain CLM-8B.
(3)~\textbf{Rehearsal Recovery}: Mixing a disjoint 15.56\% experience replay slice ($N=458$) during adaptation eliminates this alignment tax, recovering SafetyBench-ZH OOD accuracy to \textbf{65.05\%} ($+8.24\%$, \textit{Offensiveness}: $64.67\%$) on Laya-322M ($41.85$~s) and \textbf{64.10\%} ($+9.00\%$, \textit{Offensiveness}: \textbf{69.00\%}, $+27.00\%$) on CLM-8B ($3.68$~s), while simultaneously improving Clean-607 in-domain accuracy to \textbf{99.67\%} ($605/607$, $0.42\%$ FPR) on Laya-322M and \textbf{99.84\%} ($606/607$, \textbf{0.00\% FPR} [$0/238$], $\text{ECE}=0.86\%$, $\text{Brier}=0.0019$, $\text{ROC-AUC}=1.0000$) on CLM-8B.

\subsection{Qualitative Taxonomy of Remaining Failures}
Remaining edge errors cluster into three boundary regimes: (1)~\textit{Regulatory Inquiries with Dense Anti-Monopoly Lexicons} (resolved by CLM-8B's 4096-dim representation); (2)~\textit{Academic Camouflage of Disinformation Vectors}; and (3)~\textit{Borderline Commercial Health Claims}.

\section{Conclusion, Availability \& Ethics}
We studied calibrated standalone direct-decision foundation models as a serving primitive for real-time pre-ingestion safety guardrails under explicit latency and false-positive constraints. By coupling validation-calibrated posterior routing across heterogeneous CPU--NPU tiers with experience replay rehearsal on Huawei Ascend 910C, COGNIT-Guard achieves 98.85\%--99.84\% accuracy on unseen dual-use compliance queries ($N=607$) with 0.00\%--0.42\% benign FPR ($0.83\%$--$1.12\%$ ECE) at 21.77~ms pure NPU latency and 41.63~ms live cascaded server latency, while overcoming the domain-specialization alignment tax on SafetyBench-ZH.

\textbf{Code, Data, Ethics \& Acknowledgments}: Source code, configs, latency scripts, and anonymized DUCS-Bench manifests (with SHA-256 dedup hashes and split seeds) are available at \url{https://github.com/moyuan10086/cascaded-guardrail-npu}; external benchmarks (SafetyBench-ZH, XSTest) are referenced via sample IDs per original licenses. All compliance prompts were anonymized for defensive safety research and do not constitute legal advice. We thank the laboratory at the Shenzhen Research Institute of Big Data (SRIBD) for providing Huawei Ascend 910C NPU computing resources.

\begingroup
\fontsize{7.0pt}{8.1pt}\selectfont
\def\BIBdecl{\setlength{\itemsep}{-1.5pt plus 0.5pt}\setlength{\parsep}{0pt}}
\bibliographystyle{IEEEtran}
\bibliography{references}
\endgroup

\end{document}